\documentclass[aps,prb,twocolumn,superscriptaddress,longbibliography]{revtex4-1}

\usepackage{hyperref} 
\usepackage{graphicx}
\usepackage{amsmath}
\usepackage{amssymb} 
\usepackage{hyperref}
\usepackage[dvipsnames,svgnames,x11names,hyperref]{xcolor}
\hypersetup{colorlinks=true,linkcolor=NavyBlue,urlcolor=NavyBlue,  citecolor=NavyBlue}
\DeclareGraphicsExtensions{.png,.jpg,.eps}
\usepackage{xcolor}

\renewcommand{\H} {\mathcal{H}}
\newcommand{\zt} {Z_3}
\newcommand{\fz} {f}

\newcommand{\omegaz} {z}
\begin{document}

\author{Vilja Kaskela}
\affiliation{Department of Applied Physics, Aalto University, 00076 Aalto, Espoo, Finland}

\author{J. L. Lado}
\affiliation{Department of Applied Physics, Aalto University, 00076 Aalto, Espoo, Finland}

\title{Dynamical topological excitations in parafermion chains
}

\begin{abstract}
Parafermions are elusive fractional
excitations potentially emerging in fractional quantum Hall-superconductor
junctions, and represent one of the
major milestones in fractional quantum matter.
However, generic models of parafermions are not analytically solvable,
and understanding their topological modes is a bigger
challenge than conventional Majorana modes.
Here, by using a combination of tensor network and
kernel polynomial techniques, we demonstrate the emergence
of zero modes and finite energy excitations in
many-body parafermion chains.
We show the appearance of zero-energy
modes in the many-body spectral function at the edge of a topological parafermion chain, 
their relation with the topological degeneracy of the system, and
we compare their physics with
the Majorana bound states of topological superconductors. We demonstrate the
robustness of parafermion topological modes with respect
to a variety of perturbations, and we show how 
weakly coupled parafermion chains give rise to in-gap excitations.
Our results exemplify the versatility of tensor network
methods for studying dynamical excitations of interacting parafermion chains,
and highlight the robustness of topological
modes in parafermion models.
\end{abstract}

\date{\today}

\maketitle

\section{Introduction}
Unconventional excitations in quantum materials are
a central research area in modern condensed matter
physics.\cite{RevModPhys.82.3045,RevModPhys.83.1057} 
Paradigmatic examples
of unconventional excitations
are the edge modes of topological insulators,\cite{RevModPhys.82.3045,PhysRevB.78.195125,RevModPhys.88.035005} including
quantum anomalous Hall insulators and quantum spin Hall insulators.\cite{PhysRevLett.95.226801} Solely,
these systems have attracted a great amount of attention for their potential
for dissipationless electronics and spintronics. Topological superconductors\cite{Sato2017,Beenakker2013}
represent another instance in which topological excitations have a major role. 
In particular, the emergence of Majorana zero modes\cite{Majorana1937,Beenakker2013} in these systems
puts forward the possibility of using superconductors as a noise-resilient
platform for topological quantum computing.\cite{RevModPhys.80.1083,Alicea2011,PhysRevX.6.031016}

\begin{figure}[t!]
\centering
    \includegraphics[width=\columnwidth]{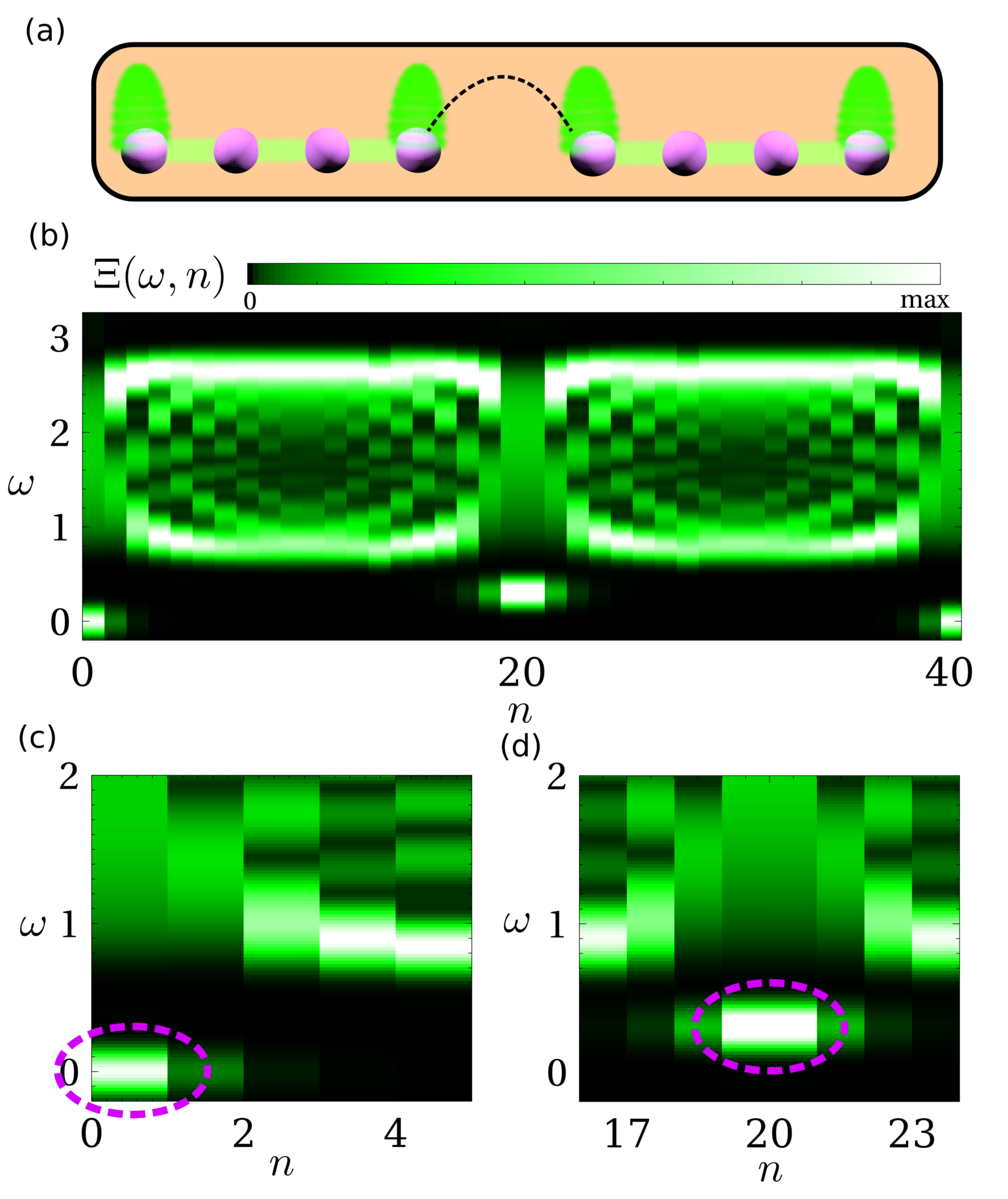}
	\caption{Sketch of two weakly coupled parafermion chains
(a), showing the emergence of decoupled topological excitations at the edge and coupled modes at the interface. 
Panel (b) shows the dynamical spectral function at each site of the
chain, showing the emergence of topological zero modes at the edges (c),
and finite energy in-gap excitations at the interface (d).
Bound states in (c,d) are highlighted by the purple dashed circles.
}
\label{fig:sketch}
\end{figure}

Interest in topological superconductors started with the first proposals
to realize artificial $p$-wave superconducting in a variety of platforms \cite{PhysRevLett.106.127001,PhysRevLett.105.177002,PhysRevB.84.144522,PhysRevLett.114.236803,PhysRevX.5.041042,PhysRevLett.111.186805,PhysRevLett.104.040502,PhysRevLett.121.037002,PhysRevLett.100.096407,PhysRevLett.106.220402}, by
combining strong-spin
orbit coupling effects, superconducting proximity
effect and exchange fields.
Majorana bound states can be generalized
to a wider class of topological
excitations, known as parafermions.\cite{Fendley2012} 
In particular, parafermions realize
quantum excitations with generalized commutation relations,
providing a powerful platform for topological quantum
computing, overcoming a limitation
of Majorana bound
states.\cite{Kitaev2003,RevModPhys.80.1083,Alicea2011}
In contrast to fermions, parafermions do not
exist in nature, and thus they must be
artificially engineered.
Following the previous point,
a variety of proposals involving
fractional quantum Hall states with superconductivity
have been put forward for their artificial
engineering,\cite{PhysRevLett.63.199,PhysRevB.59.8084,PhysRevResearch.2.023296,PhysRevLett.115.126805,PhysRevX.4.031009,PhysRevB.81.155302,PhysRevB.99.241108,PhysRevResearch.2.013232,PhysRevB.81.045323,PhysRevLett.116.216802,Moore1991,PhysRevB.100.205424,Alicea2016} inspired by the success of proposals
for Majorana bound states.

Majorana bound states can be described
in an effectively single-particle picture
with the Bogoliuvov-de-Gennes formalism.
The robustness of these modes to perturbations
stems from their topological
origin, which is associated
to the existence 
of a single-particle
non-trivial topological invariants.
Interestingly, the inclusion of many-body
interactions turns those topological models
much richer.\cite{PhysRevB.83.075103}
Ultimately,
parafermion models represent a much bigger challenge
from the theoretical point of view. 
This stems from the fact that parafermion models
cannot be solved analytically in general,\cite{PhysRevB.97.245144,PhysRevB.71.045110}
and no general proof
exists to demonstrate
ground state degeneracies.
As a result,
models for parafermions become full-fledged many-body problems,
requiring a full many-body treatment. 
Parafermion models are
substantially less explored than their Majorana
counterparts. In particular,
the computation of dynamical excitations in parafermion chains
remains a challenging problem
due to the genuine many-body nature of the problem,
and the lack of exact analytical tools for its generic treatment.

Here,
using a combination
of tensor network and
kernel polynomial techniques,
we show the emergence of edge and interface
topological excitations in parafermion chains.
In particular, here we demonstrate
that parafermion chains show edge zero modes that are resilient
to a variety of parafermion many-body interactions, and that weak coupling between
parafermion zero modes give rise to in-gap excitations at finite energy at the
interface (Fig. \ref{fig:sketch}). Furthermore, we
compare the phenomenology of these parafermion chains
with those of Majorana excitations in
topological superconductors.
Our manuscript is organized as follows. First, in section \ref{sec:clock} we present
a generalized parafermion model, 
that simultaneously captures conventional
fermions and parafermions, together with a
quantum many-body procedure used to solve the system.
In section \ref{sec:majo} we use this formalism to 
study the dynamical topological modes in a
Majorana chain, including the effects of many-body interactions,
decoupling and disorder. In Sec. \ref{sec:z3} we show that a
parafermion
chain show emergent zero-modes in the spectral function.
In section \ref{sec:z3pert} we show that perturbations
to the parafermion Hamiltonian leave the zero-edge excitation unaffected. 
In section \ref{sec:z3dec} we show how interfacial
in-gap excitations at finite energy emerge 
at the
interface between two parafermion
chains.
Finally, in section \ref{sec:con} we summarize our conclusions.

\section{Model}
\label{sec:clock}

\subsection{Clock model and parafermions}

In the following we will study
a one-dimensional model of parafermions\cite{Alicea2016,PhysRevB.91.115133,PhysRevLett.112.246403,PhysRevLett.113.066401,PhysRevB.90.165106,PhysRevB.95.075111,PhysRevB.94.125103,Fendley2012,PhysRevB.95.235127,PhysRevA.98.023614,PhysRevB.98.085143,PhysRevB.99.085113,PhysRevLett.118.170402,PhysRevLett.118.136801,PhysRevLett.116.106405,PhysRevB.99.085113}
exhibiting topological zero modes.   Parafermions are generalizations of conventional fermions with $Z_N$ symmetry showing
generalized commutation relations. Parafermion models are conveniently
written from a so-called clock model, involving operators $\tau$ and $\sigma$.
The clock operators $\tau$ and $\sigma$ generalize the Pauli $x$- and $z$-matrices, with
the following properties

\begin{equation}\label{cl}
\begin{split}
    \sigma^n = \tau^n = 1 \\
    \sigma^{\dagger} = \sigma^{n-1}\\
    \tau^{\dagger} = \tau^{n-1} \\
\end{split}
\end{equation}

The integer $N$ determines the type of $Z_N$ parafermion
considered. In particular, for $N=2$, the conventional
algebra of Pauli matrices is recovered, yielding $\sigma^2= 1$.
In contrast, for $N = 3$, one recovers the same state upon applying
the operator three times. The clock operators allow generalizing the notion of fermions, by promoting the typical Jordan-Wigner 
algebra to $Z_N$ symmetry. The clock operators follow a generalized commutation relation of the form
\begin{equation}
\label{eq:clock}
    \sigma\tau =  \omegaz\tau\sigma
\end{equation}
with $\omegaz=e^{2\pi i /n}$. In a parafermion chain, each site is taken to have its own set
of parafermion operators $\tau_j$ and $\sigma_j$.
With those local clock operators, 
the operators for parafermions are derived from the clock operators as\cite{Ortiz2012,Fendley2012,Fradkin1980}

\begin{equation}
    \chi_j = \Bigg(\prod_{k=1}^{j-1}\tau_k\Bigg)\sigma_j
\end{equation}
\begin{equation}
    \psi_j = \Bigg(\prod_{k=1}^{j-1}\tau_k\Bigg)\sigma_j\tau_j
\end{equation}
where $\psi$ and $\chi$ correspond to the two parafermion operators. 
The previous transformation can be understood as a generalized Jordan-Wigner transformation between
conventional spin operators and fermionic operators. \cite{Fendley2012} 
The Hamiltonian of the parafermion chain is constructed with $\psi_j$, $\chi_j$, $\psi^\dagger_j$, $\chi^\dagger_j$.
The parafermionic commutation relation follows from the commutation
relations of the clock operators and is given by
\begin{equation}
\begin{split}
    \chi_j\psi_k = \omegaz\psi_k\chi_j \\ 
        \chi_j\chi_k = \omegaz\chi_k\chi_j \\ 
          \psi_j\psi_k = \omegaz\psi_k\psi_j \\ 
\end{split}
\end{equation}
for $j<k$. 
These commutation relations are responsible for the exotic quantum statistics of
the chain.
In particular, taking $n=2$ recovers
the commutation relations for fermions.
Given the previous
operators, a many-body Hamiltonian
for the parafermion chain can be written as

\begin{equation}
    \H = 
    i \fz \sum_n \chi^\dagger_n \psi_n + 
    i \theta \sum_n \psi^\dagger_n \chi_{n+1} +
    \text{h.c.}
    \label{eq:h}
\end{equation}
where $\fz$ is an on-site coupling between parafermions on the same site, and $\theta$ a coupling between parafermions in different sites.
We focus on the Hamiltonian of Eq. \label{eq:h} for simplicity,
yet of course more complex Hamiltonians involving parafermion
operators can be written.
The previous Hamiltonian is known to have a rich phase diagram
for complex values of $\fz$ and $\theta$,\cite{PhysRevB.92.035154,PhysRevB.98.075421,PhysRevB.94.125103} which in particular
hosts a phase with many-body topological order. 
We note that the topological order we study
exists only in parafermionic model but not in the clock model.
In particular, in the clock chain the degenerate ground states
are not robust to local perturbations.
Here we will focus
on this topological phase, which is obtained in particular
by taking $\theta=1$ and $\fz=0.5$. In particular, we will be interested
in studying the dynamical excitations of the system, which
follows from computing the following dynamical correlator

\begin{equation}
\Xi (\omega,n)=\langle GS|\chi^\dagger_n\delta(\omega - \H+E_{GS})\psi_n| GS\rangle\
\label{eq:xi}
\end{equation}
where $|GS\rangle$ is the many-body ground state of the system
and $E_{GS}$ the ground state energy. Due to the genuine many-body nature
of this model, we will compute this dynamical correlator numerically
using the kernel polynomial tensor network as elaborated in the next section.

\subsection{Kernel polynomial tensor network formalism}
\label{sec:dmrg}

Due to the many-body nature of the Hamiltonian Eq. \ref{eq:h},
a generic analytic solution cannot, in general, be obtained. To tackle this
problem, we will here employ the tensor network
formalism\cite{White1992,PhysRevB.48.10345,Schollwck2011,ITensor,dmrgpy,2020arXiv200714822F}
which
is in particular well suited for generic interacting one dimensional
problems. In order to compute the dynamical correlators 
we will use the tensor network kernel polynomial
formalism.\cite{RevModPhys.78.275,PhysRevB.91.115144,PhysRevB.97.075111,PhysRevResearch.2.023347,PhysRevResearch.1.033009,2020arXiv200807990R}
The kernel polynomial method\cite{RevModPhys.78.275} (KPM) allows for the
computation of spectral functions directly in frequency space by performing
expansion in terms of Chebyshev polynomials
of Eq. \ref{eq:xi}.
For simplicity, we focus our discussion on
the rescaled Hamiltonian $\H \rightarrow \bar \H$, 
whose ground state energy
is located at $E=0$ and whose
excited states are restricted to the interval $[0,1)$,
\footnote{
The MPS-KPM algorithm can be performed
with a Hamiltonian whose full-spectrum is scaled and shifted
to fit the interval $(-1,1)$, which
for computational efficiency should
be done in with the spectral center
located at 0.
}
which can be generically obtained by shifting and rescaling the original Hamiltonian $\H$. The
dynamical correlator $\Xi$ for the original Hamiltonian $\H$
can then be recovered by rescaling back the energies in the
dynamical correlator $\bar \Xi$ of the scaled Hamiltonian $\bar \H$.
To
compute the dynamical correlator $\bar \Xi$, we perform an expansion of the form
\begin{equation}
\bar \Xi(\omega)=\frac{1}{\pi\sqrt{1-\omega^{2}}}\left(\mu_{0}+2\sum_{l=1}^{N_P}\mu_{l}T_{l}(\omega)\right)\
\label{KPM}
\end{equation}
where $T_l$ are Chebyshev polynomials and
$N_P$ are the number
of polynomials considered. 
The coefficients of the expansion $\mu_l$
can be then computed as
$
\mu_{l}=\langle GS|\chi^\dagger_n T_{l}(\bar \H)\psi_n|GS\rangle\
$,
where $|GS\rangle$ is computed with the
density-matrix renormalization
group (DMRG) algorithm.\cite{White1992}
Taking into account the recursion relation of the Chebyshev polynomials
$
T_l (\omega) = 2 \omega T_{l-1} (\omega) - T_{l-2} (\omega)\,,
$,
with $T_{1}(\omega) = \omega$ and $T_0 (\omega) = 1$,
the different coefficients $\mu_l$ can be computed
by iteratively defining the vectors
\begin{eqnarray}
|{w_{0}}\rangle & = & \psi_n|GS\rangle\\
|w_{1}\rangle & = & \bar {\H}|w_{0} \rangle\\
|w_{l+1}\rangle & = & 2\bar {\H}|w_{l}\rangle-|w_{l-1}\rangle\,
\end{eqnarray}
so that $|w_{l}\rangle=T_{l}(\bar \H)\psi_n|GS\rangle$.

In this way, the coefficients
$\mu_l$ are computed as
$
\mu_{l} = \langle GS |\chi_n | w_l \rangle\,.
$
To improve the convergence rate of the expansion, we
perform an autoregressive extrapolation\cite{Akaike1969} and we
quench the Gibbs oscillations
with the 
Jackson kernel.\cite{Jackson1912}

\section{Dynamical excitations in an interacting topological
superconductor}
\label{sec:majo}

\subsection{Zero modes in interacting topological superconductors}

\begin{figure}[t!]
\centering

    \includegraphics[width=\columnwidth]{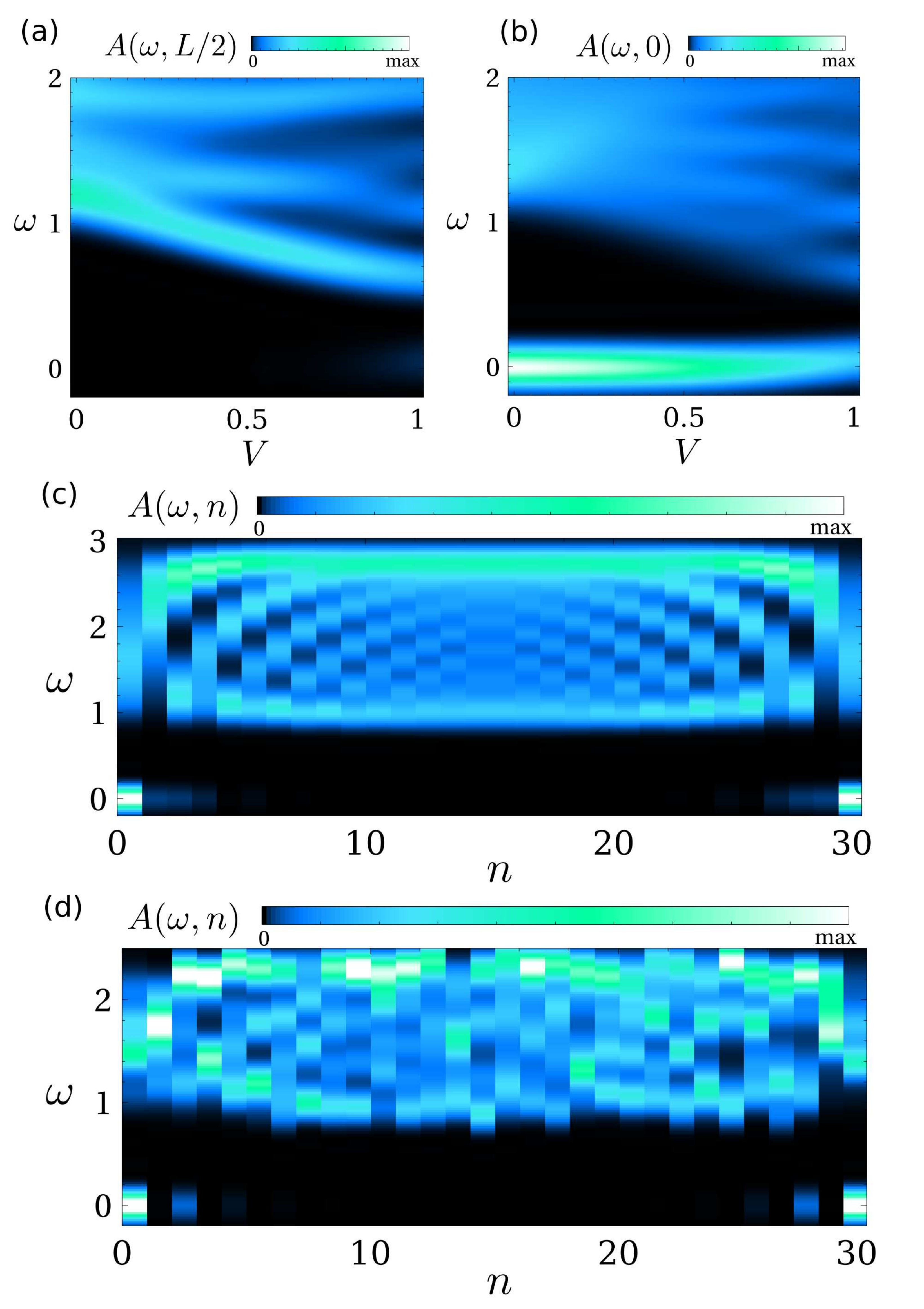}

	\caption{
	Spectral function in the bulk (a) and 
	at the edge (b) of a topological superconductor
	as a function of the interaction strength $V$.
	Panel (c) shows the spectral function in the different
	sites of the chain for the interacting topological
	superconductor, showing the presence of zero-energy modes.
	Panel (d) 
    shows the spectral function in the different
	sites of the chain including both interactions and 
	disorder, highlighting the
	robustness of the zero-energy excitations.
}
\label{fig:fig2}
\label{fig:majo}
\end{figure}

In this section, we first show how the previous formalism allows capturing
the robustness of Majorana zero modes,\cite{Kitaev2001} a well-studied
topological state that emerges taking $n=2$ in the
generalized parafermion model.
To go beyond the single-particle Majorana limit,
we will benchmark our tensor network formalism
with an interacting topological superconductor.
In particular, it is well known that the ground state degeneracy of a finite island is not lifted by many-body interactions,\cite{PhysRevLett.109.146403,PhysRevB.84.014503} and that the zero-bias
peak structure survives.\cite{PhysRevB.88.161103} 
We take the following many-body Hamiltonian for
an interacting for a one-dimensional
topological superconductor
\begin{equation}
	\begin{split}
    \H = 
     \mu \sum_n c^\dagger_n c_n + 
     t\sum_n c^\dagger_n c_{n+1} + 
    \Delta \sum_n c_n c_{n+1}   \\
    +
   V \sum_n 
   \left ( c^\dagger_n c_n -\frac{1}{2} \right )
   \left ( c^\dagger_{n+1} c_{n+1} -\frac{1}{2} \right ) +\text{h.c.}
    	\end{split}
    	\label{eq:hmajo}
\end{equation}
where $c^\dagger_n,c_n$ and the creation and annihilation
fermionic operators, $\mu$ is the chemical potential,
$t$ the hopping, $\Delta$ the
p-wave superconducting order
and $V$ the electron-electron
interaction.
In the case of $V=0$, the previous Hamiltonian corresponds
to a non-interacting one dimensional
topological superconductor, whose eigenstates
can be solved with a conventional Bogoliubov-de Gennes
transformation.\cite{Kitaev2001} This limit of $V=0$ corresponds
to the Hamiltonian of Eq. \ref{eq:h} when taking $Z_2$
operators.
In this limit $V=0$,
the previous Hamiltonian of Eq. \ref{eq:hmajo} is known to show edge
zero-modes. In particular, those
zero modes are associated with Majorana excitation, one in each edge
of the chain,
that together encode a net two-fold
degeneracy of the ground state. 
In the non-interacting regime of $V=0$, these zero-modes
can be understood as arising
from a non-trivial topological invariant of the associated
Bogoliuvov-de-Gennes Hamiltonian.\cite{Kitaev2001,Alicea2012,Beenakker2013}

In the presence of interactions $V\ne 0$,
the conventional single-particle classification no longer holds,
and the Hamiltonian becomes purely many-body. However, it is known that interactions do not lift the two-fold
degeneracy of an open Majorana chain.\cite{PhysRevLett.109.146403,PhysRevB.84.014503}
The existence of two-fold
degeneracy is associated with the emergence of a zero-energy peak
at the edge coexisting with a gapped bulk spectra in the spectral function
\begin{equation}
A (\omega,n)=\langle GS|c_n \delta(\omega - \H + E_{GS})c^\dagger_n| GS\rangle\
\label{eq:dos}
\end{equation}
where $E_{GS}$ is the ground state energy.
This can be observed by computing the dynamical
correlator of Eq. \ref{eq:dos}
at the edge and the bulk of the sample as the interaction $V$
is turned on Fig. \ref{fig:majo}ab. In particular, for $V=0$
the gapped bulk and zero-energy peak
can be understood from the single-particle picture as
mentioned above. As the interaction $V$ is increased, a
finite gap remains in bulk (Fig. \ref{fig:majo}a),
and the zero-energy peak remains (Fig. \ref{fig:majo}b).
At large enough interaction strengths, the bulk gap
would close, and the zero-energy peak would get
mixed the bulk states.
The previous phenomenology shows that, as long as interactions
are not strong enough to close the
bulk gap, the Majorana zero-energy edge mode is robust.
This can also be observed by computing the spectral function
of Eq. \ref{eq:dos} in the different sites of the chain
at an intermediate interaction $V$ as shown in Fig. \ref{fig:majo}c.
In particular, it is clearly observed that the zero-energy
modes are strongly located at the edge and that they
rapidly decay
inside the chain, leading to a gapped bulk spectra (Fig. \ref{fig:majo}c).

In the discussion above, we have considered an interacting
Hamiltonian, whose terms are uniform in space. It is, however, worth to note
that these topological zero-energy modes remain robust in the presence of 
disorder in the Hamiltonian, both in the non-interacting
and in the interacting regime. This can be explicitly shown 
by adding a disorder term to the Hamiltonian
of Eq. \ref{eq:hmajo} of the form 
$
    \H_d = \sum_n \epsilon_n c^\dagger_n c_n
$
where $\epsilon_n$ is a different random number for each site in the interval $(-\epsilon,\epsilon)$.
As shown in Fig. \ref{fig:majo}d, the edge zero modes survive in the
presence of this random disorder and interactions, 
whereas the gapped bulk states are
heavily affected by it. This resilience of the zero modes is associated with their
topological nature, signaling that for a moderate disorder
strengths the topological degeneracy of the ground state remains invariant.

\begin{figure}[t!]
\centering

    \includegraphics[width=\columnwidth]{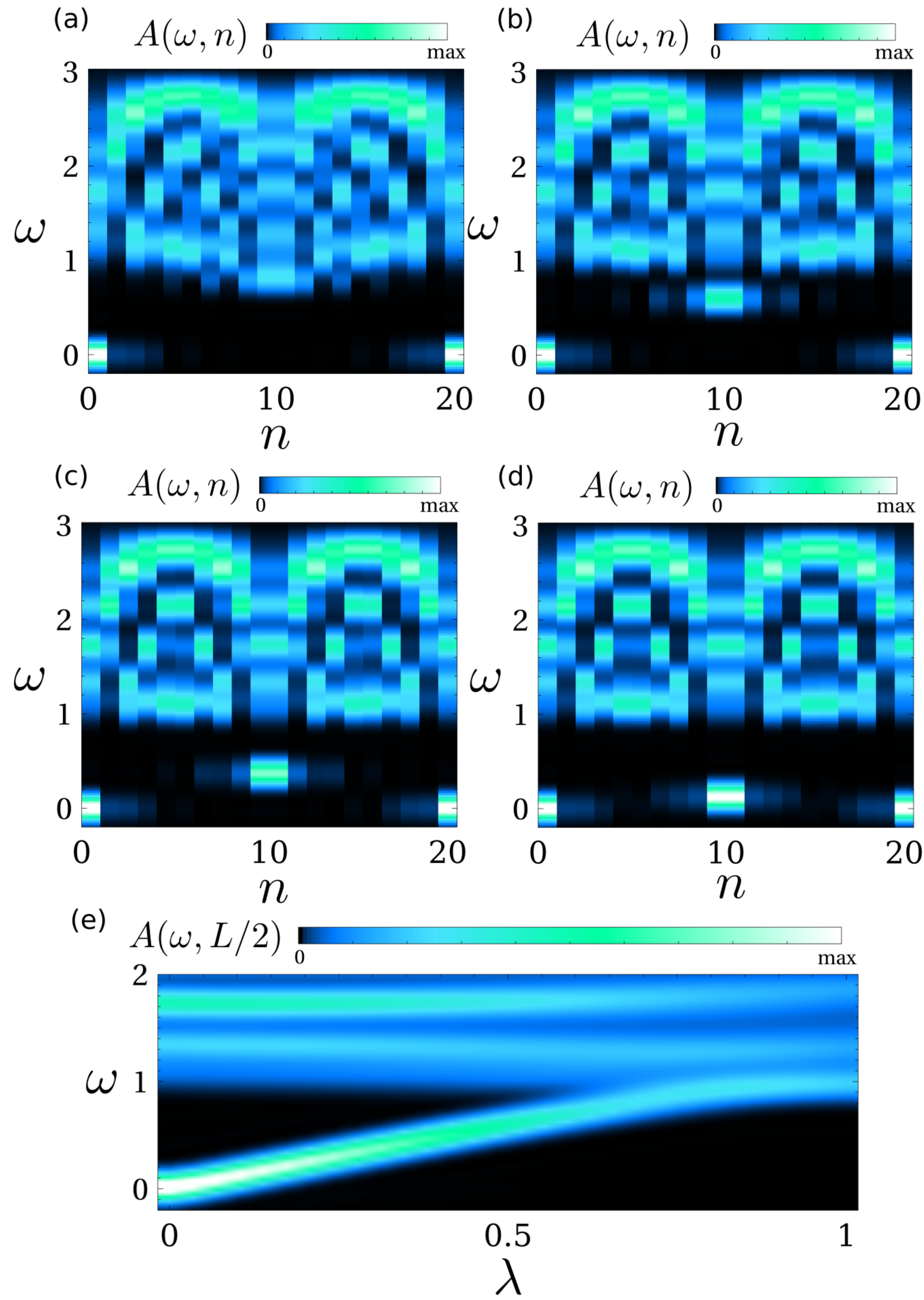}

	\caption{ (a-d) Spectral function of Eq. \ref{eq:dos} 
	in an interacting topological superconductor
	as a function of the coupling $\lambda$
	between the left and right parts,
	with $\lambda=0.7$ for (a),
	$\lambda=0.5$ for (b),
	$\lambda=0.3$ for (c),
	and $\lambda=0.1$ for (d). 
As the coupling $\lambda$ becomes weaker, interface
	modes move to lower energies,
	eventually giving rise to zero-energy excitations
	in the decoupled chain.
	Panel (e) shows the spectral function
	at the edge as a function of the decoupling,
	highlighting the emergence of the interfacial zero mode
	at $\lambda=0$.
	We took $t=1$, $\Delta=0.6$, $\mu=-0.3$, $V=0.3$
	for (a-e).
}
\label{fig:fig3}
\label{fig:majodec}
\end{figure}

\subsection{Interface excitations in coupled topological superconductors}
Previously we showed that
the topological zero-modes appear at the edge of the
one-dimensional chain, both in the presence of electronic interactions and disorder. 
We will now address how these topological zero-modes would emerge
a single chain is decoupled into two, 
which will lead to 
edge modes at each end of each subsystem.
For this goal, we now define
a parametric Hamiltonian, in which the
coupling between the left and right
parts is controlled by $\lambda$.

\begin{equation}
	\begin{split}
    \H_\lambda = & \\
     \mu \sum_{n} c^\dagger_n c_n + 
     t\sum_{n\ne L/2} c^\dagger_n c_{n+1} + 
    \Delta \sum_{n\ne L/2} c_n c_{n+1}   \\
    +
   V \sum_{n\ne L/2} 
   \left ( c^\dagger_n c_n -\frac{1}{2} \right )
   \left ( c^\dagger_{n+1} c_{n+1} -\frac{1}{2} \right ) \\
   +
    \lambda \left [ t c^\dagger_{L/2} c_{L/2+1} + 
    \Delta c_{L/2} c_{L/2+1}  \right ] \\
    +
    \lambda \left [
   V 
   \left ( c^\dagger_{L/2} c_{L/2} -\frac{1}{2} \right )
   \left ( c^\dagger_{L/2+1} c_{L/2+1} -\frac{1}{2} \right ) \right ] \\ +\text{h.c.}
    	\end{split}
\end{equation}

By definition, $\lambda=1$ corresponds to the pristine limit of Eq. \ref{eq:hmajo},
whereas $\lambda=0$ corresponds to the fully decoupled limit in which
the system consists of two independent chains.
In this limit, the Hamiltonian
consists of two fully-decoupled
chains, and therefore each chain develops its own 
pair of Majorana edge modes.
The evolution from the fully coupled
to the fully decoupled limit can be
systematically explored by computing the
the spectral
function in the chain for different strengths of the coupling $\lambda$
as shown in Fig. \ref{fig:majodec}abcd.
As it is obvious to expect, coupling
the two chains will lift away the interface zero modes.
In particular, as the chains are decoupled, an interface state emerges and drifts to
lower energies (Figs. \ref{fig:majodec}abcd). 
This can be systematically studied by looking at the evolution
of the spectral function at the interface as a function of the coupling
$\lambda$, as shown in Fig. \ref{fig:majodec}e.
In particular, the zero-mode in the fully
decoupled regime becomes a finite energy excitation as the
coupling between the two chains is increased. 
Similar phenomenology is known in non-interacting Majorana
chains, highlighting that the emergence of
finite energy excitations from coupled topological zero-modes
also holds in the purely many-body regime. In the following, we will show
that an analogous phenomenology happens in interacting parafermion chains.

\begin{figure}[t!]
\centering

    \includegraphics[width=\columnwidth]{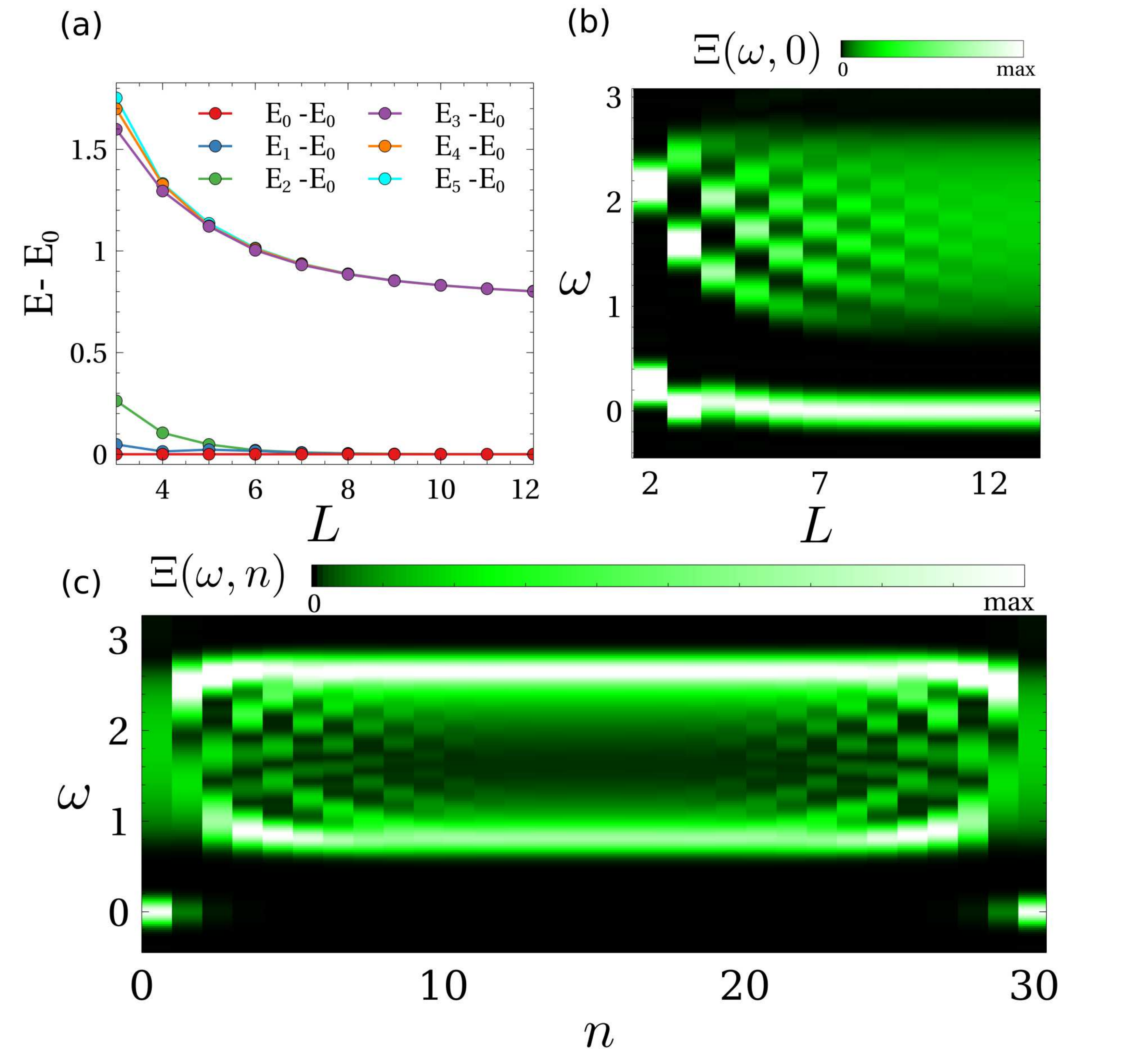}

	\caption{(a) Many-body excitation energies of a parafermion
	chain with open boundary conditions, showing the
	emergence of a three-fold degenerate ground state
	in the thermodynamic limit. Panel (b) shows the 
	spectral function at the edge as a function of the system size. 
	Panel (c) shows the
	spatially resolved spectral function, showing the
	emergence of edge excitations.
}
\label{fig:fig3}
\label{fig:para}
\end{figure}

\section{Zero-mode excitations in parafermion chains}
\label{sec:z3}
We now move on to consider chains of $Z_3$ parafermions, in particular
building on top of the previous results for an interacting
topological superconductor.
The first interesting issue to consider is the many-body degeneracy of the
parafermion chain, in comparison with the one of the
topological superconductor.\cite{Fendley2012}
This can
be observed by analyzing the excitation energies as a function of the
system size, as shown in Fig. \ref{fig:para}a. It is observed that
as the system size becomes bigger, the energies of the first two
excited states become arbitrarily
close to the ground state energy, with the next excited state presenting a finite gap.
This very same phenomenology takes
place for the Majorana model, in which
the finite splitting of the states for small chains is rationalized
in terms of the hybridization between the edge modes.
It is important to note that, in contrast with the Majorana model,
the ground state of the $Z_3$ parafermion chain becomes three-fold degenerate,
in comparison with the two-fold degeneracy of the Majorana chain.

In the case of a topological superconductor, the degeneracy of the ground state
is associated with the emergence of Majorana zero modes at the edges.
The degeneracy of the ground state with open boundary conditions for
the $\zt$ parafermion chain is again rationalized in terms
of emergent topological edge modes, but now encoding
a three-fold degeneracy. This can be observed in the
dynamical correlator computed at the edge of a parafermion chain
as a function of the chain length,
as showed in Fig. \ref{fig:para}b. In this fashion, the finite splitting
between the lowest three energy levels
for small chains can be rationalized in terms
of a finite hybridization between the topological
zero modes located at opposite edges.
Due to the existence of a finite gap in the bulk of the chain,
the zero modes are exponentially localized, leading
to an exponential dependence
of the hybridization between the states.
This can be verified by looking at the spectral function
for every site in the parafermion chain, as shown in 
Fig. \ref{fig:para}c.
In particular,
the topological zero modes are strongly
localized at edges of the chain, whereas the spectral function
remains gapped in the bulk of the chain (Fig. \ref{fig:para}c).
In the next section, we will address
the robustness of the edge zero-mode excitations,
showing that the previous phenomenology
is robust towards perturbations.

\section{Perturbations and disorder in parafermion chains}
\label{sec:z3pert}
Previously we focused on the pristine parafermionic chain
showing the emergence of topological excitations at zero-energy at the edge.
In the following, we will assess the robustness of previous zero modes with 
respect to perturbations. In particular, we will focus on two different
interaction terms, a biquadratic interaction between parafermions,
and a next to nearest-neighbor hopping in the parafermion
chain.
We will examine the impact of these perturbations by computing
the edge and bulk spectral function as the interaction term is increased,
as it was shown in the interacting topological superconductor above.

\begin{figure}[t!]
\centering

    \includegraphics[width=\columnwidth]{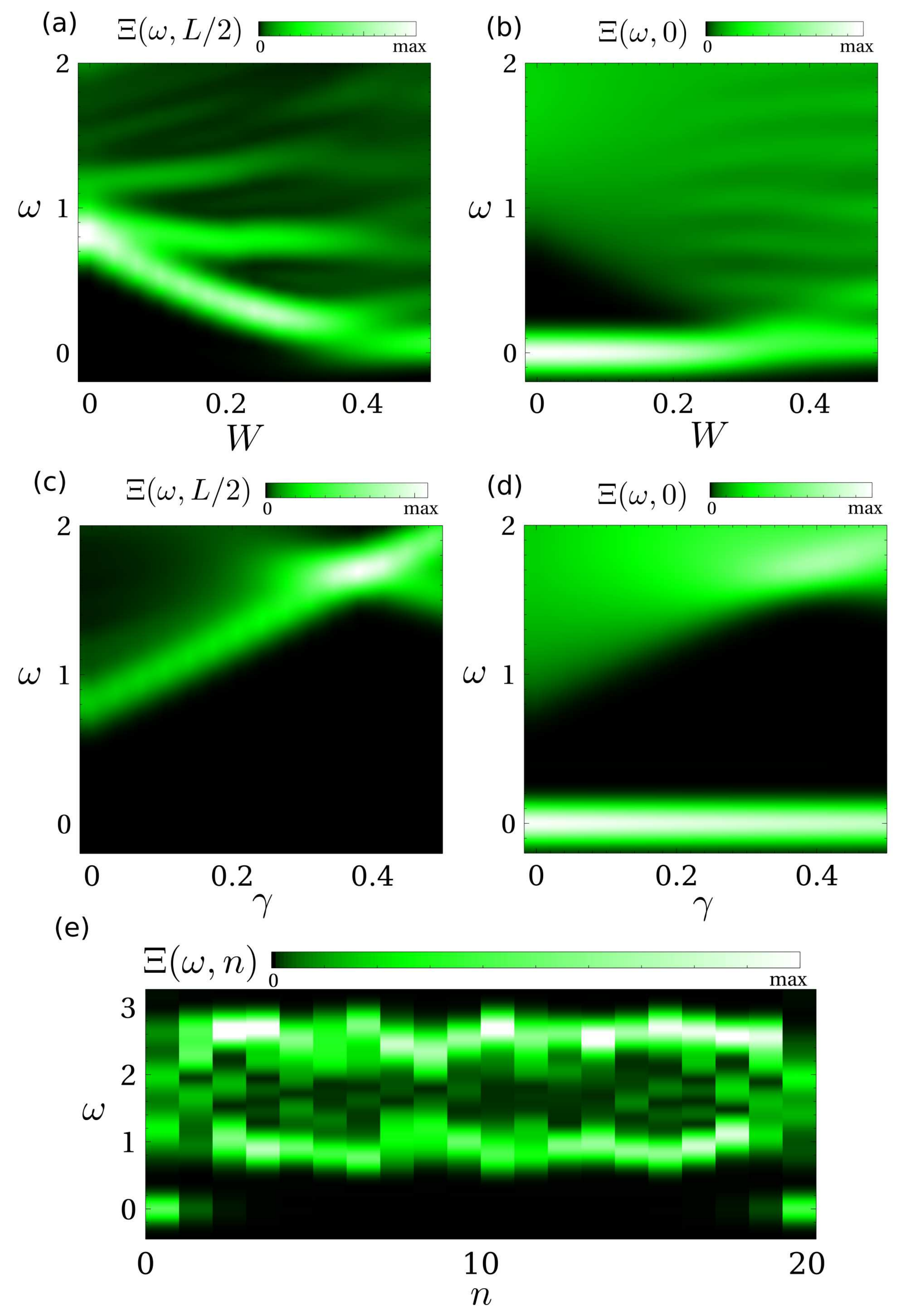}

	\caption{(a,b)
	Spectral function in the bulk (a) and at the edge (b) as
	a function of the biquadratic interaction term of Eq. \ref{eq:z3int}
	(c,d) Spectral function in the bulk (c) and at the edge (d)
	as a function of the second-neighbor hopping of Eq. \ref{eq:z3hop}
	Panel (e) shows the spectral function in every site
	for the parafermion chain with random disorder of Eq. \ref{eq:paradis}.
}
\label{fig:fig}
\label{fig:parapert}
\end{figure}

Let us first address the case of biquadratic interactions. In particular, we now
include a term in the Hamiltonian that involves four parafermionic operators,
leading to a Hamiltonian of the form
\begin{equation}
\begin{split}
    \H_W = 
    i \sum_n \fz \chi^\dagger_n \psi_n + 
    i \theta \sum_n \psi^\dagger_n \chi_{n+1} + \\
     W\sum_n \psi^\dagger_n \chi_n \psi^\dagger_{n+1} \chi_{n+1}  +   \text{h.c.}
    \end{split}
        \label{eq:z3int}
\end{equation}
where $W$ controls the strength of the biquadratic interaction.
We compute the spectral function in bulk and at the edge
as a function of the coupling parameter $W$, as shown in Fig. \ref{fig:parapert}ab.
In particular, we observe that as the interaction term is ramped up, the bulk
spectral gap decreases. However, as long as the bulk
gap remains open, the topological edge excitation
remained pinned at zero-energy. This phenomenology emphasizes that the
biquadratic interaction parametrized by $W$ competes with the topological gap.
However, as long as such perturbation is not strong enough to close the
bulk gap, the topological edge excitations will remain pinned at zero-energy.
From the point of view of the degeneracy of the ground state of the
parafermion chain, this means that a three-fold degeneracy is robust
against the biquadratic perturbation. It is interesting to note
that this is an analogous phenomenology as the one shown above for the
Majorana chain.

\begin{figure}[t!]
\centering

    \includegraphics[width=\columnwidth]{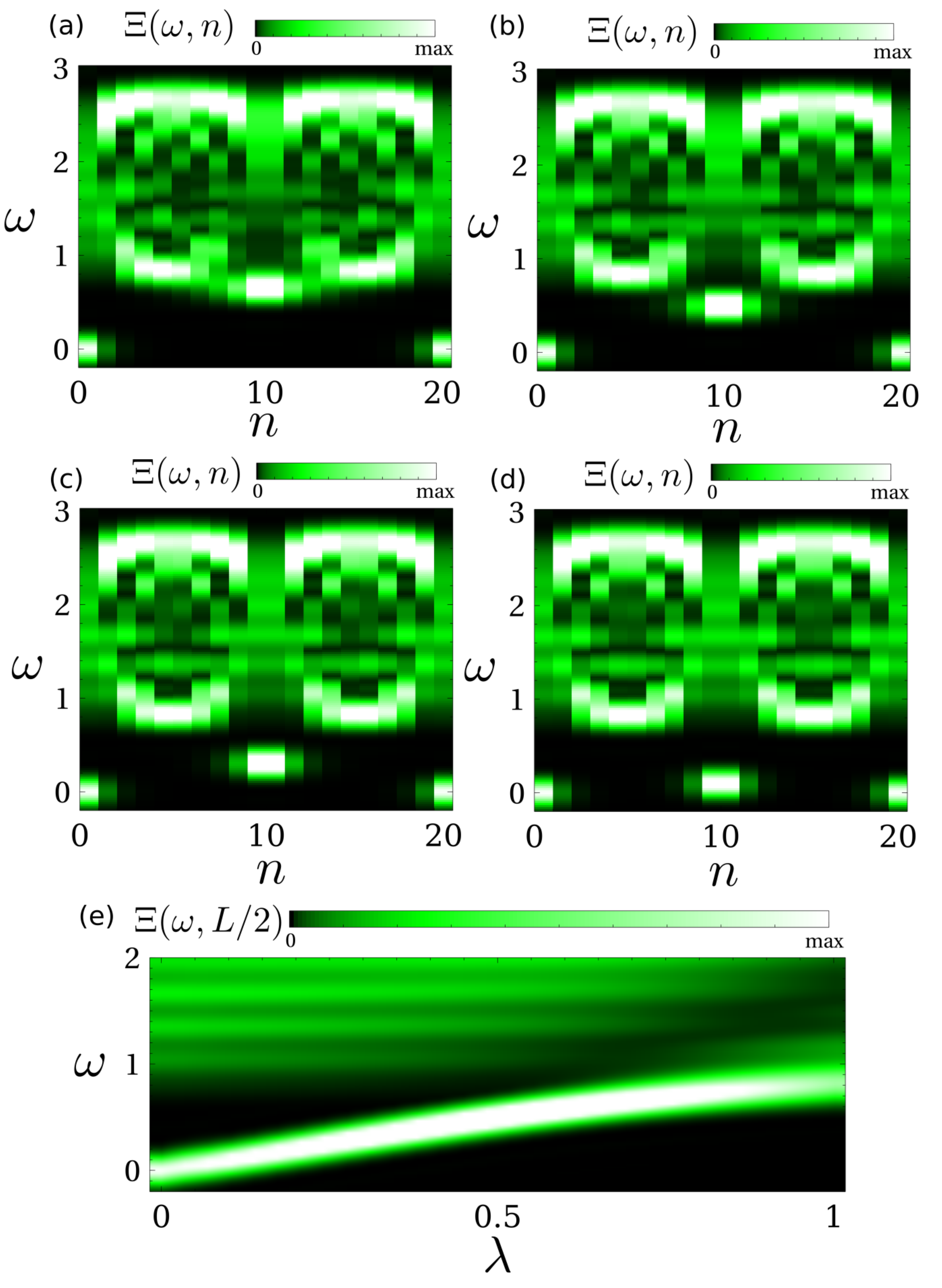}

	\caption{(a-d) Spectral function
	in the different sites
	of two coupled parafermion chains,
	for different values of the interface
	coupling $\lambda$
	between the left and right parts. We took
	$\lambda=0.7$ for (a),
	$\lambda=0.5$ for (b),
	$\lambda=0.3$ for (c),
	and $\lambda=0.1$ for (d). 
As the coupling $\lambda$ is decreased, interface
	modes move towards lower energies,
	eventually giving rise to topological
	modes in the two decoupled chains.
	Panel (e) shows the spectral function
	at the interface as a function of the coupling,
	highlighting the emergence of the interfacial zero mode
	at $\lambda=0$.
}
\label{fig:fig4}
\label{fig:paradec}
\end{figure}

After showing that first neighbor interactions compete with
the topological phase, we now turn to a different perturbation
whose effect is dramatically different.
We now consider a bilinear term in the parafermion Hamiltonian,
giving rise to a second neighbor hopping. The full Hamiltonian
now becomes
\begin{equation}
\begin{split}
    \H_\gamma = 
    i \sum_n \fz \chi^\dagger_n \psi_n + 
    i \theta \sum_n \psi^\dagger_n \chi_{n+1} + \\
    \gamma\sum_n \psi^\dagger_n \chi_{n+2}  +   \text{h.c.}
    \end{split}
    \label{eq:z3hop}
\end{equation}
where $\gamma$ parametrizes the strength 
of a
second-neighbor hopping between the parafermion
operators. We show the spectral function in the bulk as a function of the coupling
parameter $\lambda$ in Fig. \ref{fig:paradec}. In particular, we see that the spectral function
in the bulk increases its gap as $\gamma$ is ramped up. At the same time, the edge spectral function
keeps showing a zero-energy resonance corresponding to the topological
edge state. This phenomenology highlights that perturbations to the parafermionic
Hamiltonian can also enhance the topological gap, and more importantly
keeping the excitations pinned at zero-energy at the edge. 

It is finally interesting to show that the emergence of zero-modes
is not associated with the translational symmetry of the lattice.
In particular, we now consider a Hamiltonian parafermion lattice
where the couplings are disordered
\begin{equation}
    \H_{\text{dis}} = 
    i \sum_n f_n \chi^\dagger_n \psi_n + 
    i \theta \sum_n \psi^\dagger_n \chi_{n+1} +
    \text{h.c.}
    \label{eq:paradis}
\end{equation}
where now $f_n$ takes random values between $(0.37,0.62)$. We compute the spatially resolved
spectral function as shown in Fig. \ref{fig:parapert}e. It is seen that the edge zero-modes
survive in this disordered chain, despite the strong effect on the bulk states. This phenomenology
demonstrates the robustness of the zero-energy modes of the parafermion chains, and in particular, 
their existence is not associated with lattice symmetries. The next section addresses how coupling
different topological modes between different chains lifts those excitations from zero-energy.

\section{Interface excitations in coupled parafermion chains}
\label{sec:z3dec}
Previously, we showed that weak perturbations
to the parafermion Hamiltonian to not lift the edge
excitations from zero-energy.
We now explore how topological
excitations at zero-energy can be created by weakly
coupling two parafermion chains.
For this sake, we define a parametric Hamiltonian of the form

\begin{equation}
\begin{split}
\H_{\text{dis}} = 
    i f \sum_n \chi^\dagger_n \psi_n +  \\
    i \theta \sum_{n\ne L/2} \psi^\dagger_n \chi_{n+1} +
     i \lambda \theta \psi^\dagger_{L/2} \chi_{L/2+1} +
    \text{h.c.}
    \end{split}
\end{equation}
where $\lambda$ controls the coupling between two halves of the chain.
In particular, for $\lambda=1$ the system corresponds to a uniform chain, whereas for $\lambda=0$
the system is formed for two decoupled chains.

Let us now look at the spectral function at every site as a function of the
coupling strength between the two chains $\lambda$,
as shown in Fig. \ref{fig:paradec}abcd.
Like in the Majorana case, it is is obvious to expect
that coupling
the two chains will lift away the interface
zero-energy excitations.
In the pristine case, $\lambda=1$ zero-edge excitations emerge at the two-edges,
coexisting with a fully gapped bulk.
Starting with a finite but not perfect coupling $\lambda=0.7$ (Fig. \ref{fig:paradec}a),
we observe that a finite energy excitation starts to appear at the interface between the two chains.
As the
coupling between the two halves in weakened, an in-gap state drifts towards
lower energies (Fig. \ref{fig:paradec}bcd), ultimately creating zero modes
at the edges of the now two decoupled chains. This can also be systematically explored by computing the
spectral function at the interface between the two chains as a function
of $\lambda$, as shown in Fig. \ref{fig:paradec}e. It is clearly observed that the two topological edge modes,
originally located at zero-energy, become finite energy excitations as the coupling between the two chains
is increased. This shares the same phenomenology as conventional Majorana chains, highlighting that
the hybridization between topological zero modes generically give rise to finite energy excitations.
It is finally interesting to note that for $\lambda\ne 0$, the collective
ground state of the two chains will be three-fold degenerate in the
thermodynamic limit. In contrast, for $\lambda=0$ the ground state
becomes nine-fold degenerate. For $\lambda \ne 0$, the first excited state
will then correspond to the interface excitation that arises from the
coupled edge modes at the junction, whose energy can be inferred from the
spectral function of  Fig. \ref{fig:paradec}e. These results highlight that coupling topological excitations is an effective way of creating in-gap modes at finite energy.

\section{Conclusions}
\label{sec:con}
To summarize, we have shown the emergence of zero modes
and excitations at finite energies in a parafermion chain. 
Generic parafermion models are challenging
to study analytically, and 
their topological excitations 
are less understood than those of
single-particle topological
systems.
To study this interacting model,
we employed a combination of tensor network and kernel polynomial
techniques that allow addressing the full excitation spectra of the interacting Hamiltonian.
We have shown that topological parafermion chains
feature robust zero-energy excitations, 
that encode the three-fold
degeneracy of the ground state in the thermodynamic limit. 
We demonstrated that
these excitations are robust against perturbations,
including biquadratic interactions, second neighbor
hopping and disorder.
We then showed how interfacial modes at finite energies 
can be created by weakly coupling different parafermion chains,
with an excitation energy controllable
by the coupling between the chains.
Our results demonstrate
the robustness of these topological excitations
in parafermion chains, and put forward kernel
polynomial tensor networks as a versatile technique
to study finite-energy excitations
in highly interacting models.

\section*{Acknowledgments}
We acknowledge the computational resources provided by
the Aalto Science-IT project. 
J.L.L. is grateful for financial support from the
Academy of Finland Projects No. 331342 and No. 336243.

\appendix

\section{Relation between $A(\omega,n)$
and the single-particle excitation energies
}

Here we show that, in the non-interacting limit,
the spectral function of Eq. \ref{eq:dos}
corresponds to the single-particle
density of states.

Let us take a single particle Hamiltonian
of the form
$\H = \sum_{ij} H_{ij} c^\dagger_i c_j$.
In its diagonal form, it becomes
$\H = \sum_k \epsilon_k d^\dagger_k d_k$, where $d_k$
are the single particle eigenstates and $\epsilon_k$
the single particle eigenenergies.
A unitary transformation $U$ relates the $c^\dagger_n$ and $d^\dagger_k$
operators as $c^\dagger_n = \sum_{k} U_{k,n} d^\dagger_k$.
We start with the single-particle formalism to
compute the local spectral function,
also known as the local density of states.
For the single-particle formalism, we take a basis
of single-electron states $|n\rangle = c^\dagger_n |0\rangle$,
where $|0\rangle$ is the vacuum state.
In this case,
we would compute the local density of states as
$D(\omega,n) = \langle n | \delta (\omega - H) | n \rangle
\sim 
\langle n | \text{Im} [(\omega - H)^{-1}] | n \rangle
$,
where $H$ is the tight-binding matrix.
In the diagonal basis, it takes the form
$D(\omega,n) = \sum_{k} |U_{k,n}|^2 \delta (\omega - \epsilon_k)$,
which is the conventional form for the local spectral function.

We now move on to the computation in the spectral function
working in the many-body Fock space as done in DMRG.
\cite{PhysRevB.84.075139,PhysRevB.83.195115,PhysRevB.52.R9827,PhysRevB.83.161104}
In this space, the basis is no-longer states with a single electron,
but with an arbitrary number of electrons.
The many-body ground state
is given by the
Fermi sea $|GS\rangle = \Pi_{\epsilon_k<0} d^\dagger_k |0\rangle$,
where $|0\rangle$ is the vacuum state $d_k |0\rangle = 0$.
This state is, by definition, the eigenstate with the lowest possible
energy associated to the Hamiltonian $\H$, with
$\H|GS\rangle = E_{GS} |GS\rangle $,
and it has a total energy
$E_{GS}=\sum_{\epsilon_k<0} \epsilon_k $.
The excited states can be build analogously. In particular,
for each single-particle energy $\epsilon_{k'}>0$, there is
an excited many-body
state with one more electron than the ground state
given by $|k'\rangle = d^\dagger_{k'}|GS\rangle$,
with an energy $\H|k'\rangle = (E_{GS}+\epsilon_{k'}) |k'\rangle $.
With the previous points, let us now move on
to consider the dynamical correlator of Eq. \ref{eq:dos}
$A(\omega,n) =
\langle GS | c_n \delta(\omega-\H+E_{GS}) c^\dagger_n |GS\rangle$.
In the Fock space, the term $\delta(\omega-H+E_{GS})$ takes the
form $\sum_\Psi \delta(\omega-E_\Psi+E_{GS}) |\Psi\rangle \langle \Psi|$,
where $E_\Psi$ is the many-body energy of the many-body eigenstate
$|\Psi \rangle$. With the previous representation
and taking the definitions of the different terms, we get
$A(\omega,n) = \sum_{\epsilon_k>0} |U_{k,n}|^2 \delta(\omega-\epsilon_k)$,
the conventional definition of the local density of states for
a single particle Hamiltonian.
In other words, the function $A(\omega,n)$
directly reflects the single particle energies
$\epsilon_k>0$, namely the unoccupied single particle states.
As a result, the spectral function computed with Eq. \ref{eq:dos}
corresponds to the conventional single-particle
spectral function as $D(\omega>0,n) = A(\omega,n)$.
We note that the full local density of states
can be computed analogously as
$D(\omega,n) =
\langle GS | c_n \delta(\omega-\H+E_{GS}) c^\dagger_n |GS\rangle
+
\langle GS | c^\dagger_n \delta(-\omega-\H+E_{GS}) c_n |GS\rangle
$. For an interacting system that lacks a single particle
description, the previous formalism allows computing the
spectral function generically, and therefore
it is the method used in our manuscript.

\bibliography{biblio}{}

\end{document}